\documentstyle[12pt,a4]{article}
{\begin{quote}{#1:}}%
{\end{quote}}
\raggedright
\author{\\ \normalsize  E.~L.~Afraimovich, E.~A.~Kosogorov,
           L.~A.~Leonovich, O.~S.~Lesyuta, and I.~I.~Ushakov\\
\vspace*{0.3cm}
\\
\sl Institute of Solar-Terrestrial Physics, Russian Academy of Sciences,\\
\sl 664033~~Post Box~4026, Irkutsk, Russia, e-mail:afra@iszf.irk.ru}
\title{\vspace{2.3cm}\Large {\bf
INSTANTANEOUS RESPONSE OF THE
IONOSPHERE TO A SUDDEN COMMENCEMENT OF THE STRONG \mbox{MAGNETIC} STORM OF APRIL 6, 2000}}
\date{}
\begin{document}
\maketitle
\vspace*{0.8cm}
\begin{abstract}
\begin{flushleft}
\thispagestyle{empty}
\parindent=0cm
\vspace*{0.2cm}
%\vspace*{\baselineskip}
We developed a new technology for global detection of atmospheric
disturbances, on the basis of phase measurements of the total electron
content (TEC) using an international GPS networks. Temporal
dependencies of TEC are obtained for a set of spaced receivers of the
GPS network simultaneously for the entire set of visible satellites.
These series are subjected to filtering in the selected range of
oscillation periods using known algorithms for spatio-temporal
analysis of signals. An "instantaneous" ionospheric response
to the sudden commencement of a strong magnetic storm of April 6, 2000 was
detected. On the dayside of the Earth the largest value of the net
response amplitude was found to be of order $0.8\times10^{16}$
m${}^{-2}$ (1--2$\%$ of the background TEC value), and the delay
with respect to the SC in mid-latitudes was about 200 s. In higher
latitudes the delay goes as long as 15 min. On the nightside these
values are $0.2\times10^{16}$ m${}^{-2}$  and 30 min, respectively.
The velocity of the traveling disturbance from the middle to high latitudes
on the dayside as well as from the dayside to the nightside was about
10-20 km/s.
\end{flushleft}
\end{abstract}
\vspace*{0.2cm}
\section*{\rm \normalsize INTRODUCTION}
\label{GLOB-sect-1}
Mid-latitude ionospheric effects of geomagnetic
disturbances of different origins were addressed in many
publications, including a number of thorough reviews (Hunsucker,
1982; Hocke and Schlegel, 1996). It is now a
well-established fact that the auroral zones of the northern and
southern hemispheres generate acoustic-gravity waves (AGW) and
traveling ionospheric disturbances (TIDs), caused by them,
of the type of solitary wave of about 1-hour duration propagating
in the equatorial direction with sound and subsonic speeds (from
100 to 300 m/s). The occurrence delay $\tau$ TIDs over various
observing stations in mid-latitudes reaches $10^4$ s.
The propagation velocity fundamentally distinguishes the TIDs from
sudden ionospheric \mbox{disturbances} (SIDs) which are a virtually
instantaneous ($\tau$ no more than $10^{-3}$ s) response of the dayside
ionosphere to an increase in ultraviolet radiation intensity observed
during chromospheric flares on the Sun (\mbox{Mitra}, 1974).
The physical mechanisms for generation of the above-mentioned
types of \mbox{disturbances} were discussed in a large number of theoretical
publications, and  are  considered to be relatively reliably
established.
By \mbox{analyzing} the ionospheric effects from a strong magnetic storm of
April 6, 2000, we detected, most likely, a new manifestation of geomagnetic
disturbances in the mid-latitude \mbox{ionosphere} implying an "instantaneous"
(compared to the above-mentioned TID effects) ionospheric response to
the magnetic storm sudden commencement (SC).
For detecting this effect, we used a global spatial averaging of the
variations in total electron content (TEC) obtained from the data from the
international GPS network. Currently (as of June 2000) this network
includes at least 712 points, the data from which are posted to the
INTERNET. High-precision measurements of the TEC along the line-of-sight 
(LOS) between the receiver on the
ground and transmitters on the GPS system satellites covering the
reception zone are made using \mbox{two-frequency} multichannel receivers of
the GPS system at almost any point of the globe and at any time
simultaneously at two coherently coupled frequencies
$f_1=1575{.}42$ MHz and $f_2=1227{.}60$ MHz. The sensitivity of
phase measurements in the GPS system is sufficient for detecting
irregularities with an amplitude of up to $10^{-3}$--$10^{-4}$ of the
diurnal TEC variation. This makes it possible to formulate the problem
of detecting ionospheric disturbances from different sources of artificial
and natural origins.
\section*{\rm \normalsize RESULTS OF OBSERVATIONS}
\label{GLOB-sect-2}
For the time interval 16:00-18:00 UT, Fig. 1 presents the variations in
magnetic flux (a), 5 MeV \mbox{proton} flux (b) at geostationary orbit of the
GOES10 station ($135^\circ W$), and of the H-component of the magnetic field at station
Irkutsk ($52{.}2^\circ N$; $104{.}3^\circ E$ - c) for April 6, 2000. The
time of the geomagnetic disturbance SC, determined from these data and
from the data from other ground-based magnetic observatories, including
those located in the western hemisphere (on the dayside), corresponds
to 16{:}42 (16.7) UT. The time of SC is shown on panels (a-e) by a vertical
dashed line.
The geometry of the part of the global GPS network that was used in this
study when \mbox{analyzing} the ionospheric response to the SC of the strong
magnetic storm of April 6, 2000 (180 stations), is presented in Fig. 2a.
Dots correspond to the location of GPS stations; we do not give here
their \mbox{coordinates} for reasons of space. The upper scale indicates
the local time LT, corresponding to the time 16 UT. As is evident from
Fig. 2a, our selected set of GPS stations cover reasonably \mbox{densely}
North America and Europe, and much less densely the Asian part of the
territory used in the \mbox{analysis}. An even smaller number of GPS
stations are in the Pacific and Atlantic Oceans. However, coverage of the
territory with partial LOS to the satellites for our selected limitations
to the LOS  elevations \mbox{$\theta$ > $10^\circ$} is substantially wider.
Panel b shows the coordinates of subionospheric points for the height of
the F2-layer maximum $h_{max}$ = 300 km for all satellites visible at the
SC time for each of the GPS stations marked on panel a (a total of 746 beams).
An \mbox{analysis} of the TEC data from the selected GPS stations revealed
that almost all stations over the time interval 16:30-17:30 UT, containing
the SC time, show a single negative disturbance of about 20-min
duration. Upon removing the trend and smoothing with a time window of
30 min, we were able to determine for each beam to the satellite the
amplitude A and the time $t_{min}$  at which a minimum TEC value was
attained. Fig. 3 presents the latitudinal dependencies (obtained for each
beam) of the $t_{min}$ -a) and of the amplitude A - b) of the
ionospheric response to the magnetic storm SC, as well as the
distributions of $t_{min}$  - c) and A - d) as a function of local time LT.
The SC time is shown in panels e and j by a horizontal straight line. Thick
curves show the approximating dependencies obtained from all counts using
polynomials of degree 4.
The scatter of the position of $t_{min}$  is due 
to the fact that when the trend is removed with a time window of 30 min, 
the response to the SC is always overlapped by existing TEC oscillations 
with similar periods and with a random phase. Therefore, identifying the 
response requires a coherent combination of TEC variations for all LOS. 
The result of such a global spatial averaging of dI(t) for 511 LOS on the 
dayside is shown in panel d. A similar result for 235 LOS on the nightside 
is presented in panel e.
\section*{\rm \normalsize DISCUSSION AND CONCLUSIONS}
\label{GLOB-sect-4}
An analysis of the data in Fig. 1 suggests the conclusion that the
ionospheric response to the SC has the form of a single negative
disturbance of about 20-min duration. On the dayside of the Earth the
largest value of the net response amplitude was found to be of order
$0.8\times10^{16}$ m${}^{-2}$ (1--2$\%$ of the background TEC value),
and the delay with respect to the SC in mid-latitudes was about 200 s.
In higher latitudes the delay goes as long as 15 min. On the nightside
these values are $0.2\times10^{16}$ m${}^{-2}$  and 30 min, respectively.
Of special note is that the onset of negative TEC disturbance on the
dayside, marked according to the level of 0{.}5 from a maximum
deviation (panel d), coincides with the magnetic flux SC (panel a)
and is 120 s ahead of the SC time determined from the data from ground-based
magnetic observatories. The velocity of the traveling disturbance from middle to high latitudes
as well as from the dayside to the nightside is estimated at about 10-20
km/s. Thus our detected disturbance (a global delay $\tau$ no more than
$10^{2}$ - $10^{3}$ s) is inexplicable in terms of the AGW model, and it should
be sought when modeling a electromagnetic set of phenomena
accompanying a strong geomagnetic disturbance.
It is not improbable that in the analysis of the mechanism it would be
useful to take into account some important characteristics of the "global
detector" which we are using, such as primarily the sensitivity, continuity
and global character. However, it may well be of crucial importance that,
unlike conventional techniques of ionospheric observations, the altitude
limit of which does not exceed 500 km (ionosondes, HF Doppler
measurements) or 1000-2000 km (incoherent scatter radars, and stations
for recording the rotation of the polarization plane of the VHF signal
from geostationary satellites), it has become possible, for the first time,
to carry out a global detection of the Earth's plasmaspheric disturbances in
the height range as high as 20,000 km.
\section*{\rm \normalsize ACKNOWLEDGEMENTS}
\label{TSE-sect-6}
Authors are grateful to A.~V.~Mikhalev, E.~A.~Ponomarev and
~A.~V.~Tashilin for their encouraging interest in this study and
active participations in discussions. Thanks are also due
V.~G.~Mikhalkovsky for his assistance in preparing the English
version of the \TeX-manuscript. This work was done with support
under RFBR grant of leading scientific schools of the Russian
Federation No. 00-15-98509 and Russian Foundation for Basic
Research (grant 99-05-64753).

\end{document}